\newif\ifprl
\newif\ifheaders
\headerstrue

\ifprl
	\documentclass[11pt,aps,reprint,prl]{revtex4-1}
\else
	\documentclass[11pt,twoside,letterpaper,twocolumn]{article}
\fi

\pdfoutput=1
\usepackage[margin=20mm]{geometry}

\usepackage{url}
\usepackage{graphic x} 
\usepackage{amssymb,amsmath}

\ifprl
\else
	\usepackage{cite}
\fi

\begin{document}

\title{Violation of Bell's inequality in fluid mechanics}

\date{\today}

\ifprl
\else

	\author{Robert Brady and Ross Anderson\\
        \small University of Cambridge Computer Laboratory\\
        \small JJ Thomson Avenue, Cambridge CB3 0FD, United Kingdom\\
        \small \texttt{robert.brady@cl.cam.ac.uk, ross.anderson@cl.cam.ac.uk}}
	
	\twocolumn[
		\begin{@twocolumnfalse}
			\maketitle
			\begin{abstract} 
				\noindent We show that a classical fluid mechanical system can violate Bell's inequality because the fluid motion is correlated over large distances.
			\end{abstract}
		\end{@twocolumnfalse}
	]

\fi




The observed violations of Bell's inequality show that quantum mechanics cannot be modelled using local hidden variables~\cite{bell1964einstein,aspect1982experimental,giustina2013bell}. 
This has led to debate about non-local hidden variables~\cite{kochen1968hidden,leggett2003nonlocal} and about `locality' given that signals do not exceed the speed of light~\cite{deangelis2007experimental}.
Some authors suggest, more generally, that the Bell tests rule out models with only local interactions~\cite{kumar2011quantum}; but 't Hooft and Vervoort have separately advanced the possibility that Bell's inequality
may be violated in systems such as cellular automata that interact
only with their neighbours but have collective states of correlated
motion~\cite{thooft2009entangled,vervoort2013bell}, while Pusey,
Barrett and Rudolf argue that a pure quantum state corresponds
directly to reality~\cite{pusey2012reality}.

In this paper we show that Bell's inequality can be violated in a completely classical system. 
In fluid mechanics, non-local phenomena arise from local processes. 
For example, the energy and angular momentum of a vortex are delocalised in the fluid. 
Here we show that Euler's equation for a compressible inviscid fluid has quasiparticle solutions that are correlated in precisely the same way as as the quantum mechanical particles discussed in Bell's original paper. This correlation violates Bell's inequality.

\ifheaders
	\section*{Locality in fluid mechanics}
\else
	{\bf Locality in fluid mechanics.}
\fi

Collective phenomena in fluid mechanics behave locally in some respects, and non-locally in others. 
To see this, consider a vortex in a compressible inviscid fluid. 
The local aspects of the motion can be understood by treating the vortex as if it were a point in two dimensions located at its centre. 
The resulting trajectories can be complex or chaotic~\cite{aref1982integrable,eckhardt1988integrable}. 
The non-local aspects can be understood from the energy and angular momentum. 
In cylindrical coordinates $(r, \theta)$, the flow speed is given by $u = C/r$ where $C$ is the circulation. The kinetic energy is $\int \frac{1}{2} \varrho u^2 . 2 \pi r dr \approx \pi \varrho C^2 \log r$ and the angular momentum $\int \varrho u r . 2 \pi r dr \approx \pi \varrho C r^2$ 
where $\varrho$ is the density per unit area.
The lower limit of $r$ is given by effects such as cavitation, which are not of interest here. 
However, from the upper limit, the energy and angular momentum reside at large distance from the centre; furthermore they would both be unphysically large if there were no other factor~\cite{faber1995fluid}. 

The relevant factor in fluid mechanics is that vortices are created in pairs of opposite circulation, which have no net angular momentum and whose energy is finite since the fluid velocities are opposed at large distance.
This precise opposition cannot switch off in a lossless fluid. It follows that
the vortex coupling does not weaken with the distance $d$ between the centres and the fluid motion remains correlated over distances comparable to $d$. 

Bell's requirement of locality is that of Einstein, Podolsky and Rosen: `the result of a measurement on one system be unaffected by operations on a distant system with which it has interacted in the past'~\cite{einstein1935can,bell1964einstein}. 
This may seem reasonable if we consider just the vortex cores; but it is not 
clearly defined as regards the energy and angular momentum in the fluid. 
If an `operation' is to couple effectively to the delocalised energy and angular momentum, its influence must normally extend over a large enough region to affect both systems. We will now explore this in detail.
%

\begin{figure}[htb]
  \centering
	\includegraphics[width=0.25\textwidth]{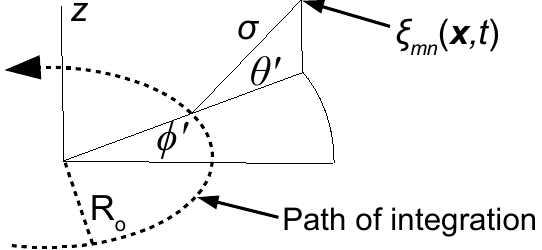}
	\caption{\em The coordinates of equation \eqref{eq:xi-mn}.}
	\label{fig:sonon-integration-path}
\end{figure}

\ifheaders
\section*{Quasiparticle solutions}
\else
	{\bf Quasiparticle solutions.}
\fi
Euler's equation for a compressible inviscid fluid has families of
quasiparticle solutions which exhibit similar behaviour in three dimensions rather than two. 
They resemble vortex rings, but the flow is irrotational everywhere. 
We will adopt the usual simplification of fluid mechanics~\cite{faber1995fluid} in which the density per unit volume $\rho({\bf x}, t)$ is extended to a complex function $\xi({\bf x}, t)$ so that $\rho = \rho_o(1+ \Re(\xi))$ where $\rho_o$ is the mean density and $\Re$ means the real part. The solutions~\cite{brady2013incompressible} are proportional to
\begin{equation}
	\xi_{mn}({\bf x}, t) = \int_0^{2 \pi} e^{-i (\omega_o t + m \theta' - n \phi')} j_m (k_r \sigma) k_r R_o d\phi'
	\label{eq:xi-mn}
\end{equation}
The coordinates are as in figure \ref{fig:sonon-integration-path}, $j_m$ is a spherical Bessel function of the
first kind
and $m$
and $n$ are integers. 
We are interested in the case $m = 0$ or $\pm 1$. The Legendre polynomial $P_m(\cos \theta')$ simplifies to  $\cos m \theta'$, so the integrand (at fixed $\phi'$) will be recognised as a spherical solution to the wave equation. Hence, like sound waves, equation \eqref{eq:xi-mn} also obeys Euler's equation for a compressible fluid at low amplitude~\cite{faber1995fluid}.
The negative mean Bernoulli pressure near
the centre perturbs the solution, reducing $R_o$ to a small size which is calculated in~\cite{brady2013incompressible}.

\begin{figure}[htb]
	\centering
		\includegraphics[width=0.45\textwidth]{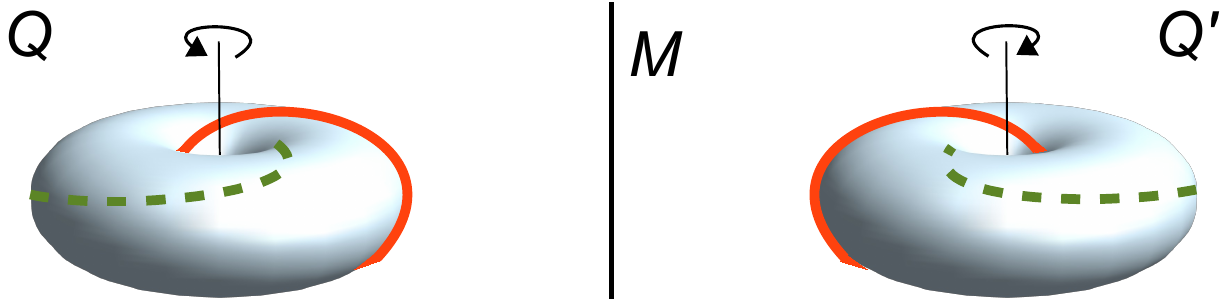}	
		\caption{\em Near-field sketch of the compressions (full line) and rarefactions (dotted) of the $\xi_{z+}$ quasiparticle ($Q$) and $-\xi'_{z-}$ ($Q'$). The far field oscillations are discussed in the text.}
	\label{fig:aligned-sonons}
\end{figure}

The quasiparticle $\xi_{11}({\bf x}, t)$ in \eqref{eq:xi-mn} is
sketched as $Q$ in figure \ref{fig:aligned-sonons}. It has a line of
compression which circles both diameters of the ring and, from the
factor $e^{-i(\omega_o t - \phi')}$, it rotates in the conventional $+z$
direction.  It will be called, more simply, $\xi_{z+}$.

Its mirror image, $Q'$, is $-\xi_{1 -1}({\bf x - x'}, t)$ where $\bf
x'$ is the new position. It has the opposite chirality and rotates in
the opposite direction, and will be called $-\xi'_{z-}$.  It is used
in the following analysis but $-\xi_{-1-1}({\bf x - x'}, t)$ is
similar.

Just as vortices bind in pairs of opposite circulation, so quasiparticles bind in pairs rotating in opposite directions.
This can be seen by considering the waves propagating
around the $z$ axis in figure \ref{fig:aligned-sonons}, which carry momentum
with them.  From Euler's equation at low amplitude, the fluid speed is
$u_\phi = \pm c \Delta \rho/\rho_o$ where $c$ is the speed of sound
and the sign depends on the direction of propagation~\cite{faber1995fluid}.
The angular momentum of a fluid element is $r_z u_\phi (\rho_o +
\Delta \rho) dx^3$ where $r_z$ is measured from the $z$ axis, so the total is
$\pm c \int r_z (\Delta \rho)^2 dx^3 / \rho_o$. Like a vortex, this grows without limit at large radius, when $\Delta \rho \propto \sin(r+P)/r$
thanks to the Bessel function in \eqref{eq:xi-mn}, where
the phase $P$ and the amplitude depend on the spherical angle. Since angular momentum is finite and conserved, the
quasiparticles must always be in pairs rotating in opposite directions (or larger groups with finite angular momentum) even if they are arbitrarily far apart.

\ifheaders
	\section*{Bragg mirror}
\else
	{\bf Bragg mirror}
\fi
The energy of a quasiparticle in its ground state does not radiate away because it is reflected by other quasiparticles. 
The perturbations to the outgoing radiation on account of interacting quasiparticles in the ground state are phase aligned as in a Bragg mirror.
This phase alignment occurs spontaneously and is observed in coupled oscillators.
Two pendulum clocks
mounted together spontaneously align antiphase, with each blocking the escape of vibrational energy from the
other~\cite{bennett2002huygens}; and larger systems have a regime where
the oscillators are correlated by an order
parameter~\cite{acebron2005kuramoto}, which equilibrates and preserves
the energy of the individual oscillators.
In the case of quasiparticles, the phase can be understood from the energy. 
If an isolated quasiparticle could be created, its kinetic energy $U = \int
\frac{1}{2} \rho u^2 . 4 \pi r^2 dr$ would diverge at large $r$
by the above substitution.
But this energy is finite if the quasiparticles interfere destructively at
large distance, which is thus their preferred alignment.

Consider the lowest-order correlation, which couples quasiparticles
into pairs. Destructive interference at large distance translates into
constructive interference in the space between a pair, since the
integrand in \eqref{eq:xi-mn} changes phase by $\pi$ compared to a
freely propagating wave when going through the centre.  Thus, the
lowest energy state has an antinode of $\Delta \rho$ on the mirror
plane $M$ and a mirror image alignment such as in figure
\ref{fig:aligned-sonons}. 

Two quasiparticles aligned in this way may be compared to two vortices of opposite circulation. In both cases the fluid velocities reinforce between them and are opposed at large distance, and in both cases the associated fluid energy increases with the separation, resulting in an attractive force if the centres are constrained not to move. The attractive force in quasiparticles is calculated in~\cite{brady2013incompressible}.

\ifheaders
	\section*{Classical superposition}
\else
	{\bf Classical superposition.}
\fi
In fluid mechanics, waves are treated as linear superpositions which are perturbed by the nonlinear terms in Euler's equation~\cite{faber1995fluid}.
Applying the same approach, a quasiparticle can be in a superposition of two components rotating in opposite directions, $\xi = \alpha_1 \xi_{z+} + \alpha_2 \xi_{z-}$. 
The $\alpha_i$ give the amplitudes and phases of the components, and so they are complex numbers, but the physical quantity $\Delta \rho \propto \Re(\xi)$ is real. 
A simple example is $\alpha_1 = \alpha_2 = 1/\sqrt2$, where the factor $e^{-i(\omega_o t - \phi')}$ in \eqref{eq:xi-mn} has been replaced by a standing wave $\sqrt2 ~ e^{-i \omega_o t} \cos \phi'$.

As is typical for ordinary waves, the perturbations are independent when assessed in terms of the energy $U \propto \int (\Delta \rho)^2 dx^3$, since the cross terms vanish due to the factor $e^{i n \phi'}$ in \eqref{eq:xi-mn}, giving $U \propto |\alpha_1|^2 + |\alpha_2|^2$.
The degenerate states of constant $U$ can be written
\begin{equation}
	(\alpha_1, \alpha_2)~~=~~ e^{i(S -\tfrac12 \varphi)}\left(\cos \tfrac12 \theta, ~e^{i \varphi} \sin \tfrac{1}{2} \theta \right)
\label{eq:bloch-state}
\end{equation}
where $\theta$, $\varphi$ and $S$ are a real numbers to be determined; note that $U$ is automatically constant and higher order perturbations have been neglected.

To interpret the quantity $\theta$, consider the scalar
\begin{equation}
	\sigma_z 
	~~=~~ \frac{|\alpha_1|^2 - |\alpha_2|^2}{|\alpha_1|^2 + |\alpha_2|^2}
	~~=~~\cos \theta
\label{eq:cos-theta}
\end{equation}
The state $(\alpha_1, \alpha_2) = (1,0)$ or $\xi_{z+}$ has all its rotational energy in the conventional $+z$ direction. 
It has $\theta = 0$ (or $2 \pi$ etc.) which we associate with $+z$. Likewise, from $(0,1)$, we associate $\theta = \pi$ with the $-z$ direction. A similar analysis applies for the directions in between, so that $\theta$ can be interpreted as a physical direction in space, and \eqref{eq:cos-theta} as the projection of the rotational energy on to the $z$ axis, normalised to a maximum value of 1. 
The same interpretation follows by noting that $|\alpha_1|^2$ is proportional to the energy of the component rotating in the $+z$ direction and $|\alpha_2|^2$ the $-z$ direction.

If this projection executes a complete rotation, so that $\theta$ increases by $2 \pi$, then \eqref{eq:bloch-state} reverses sign. This system is completely classical and it is not claimed this shows it is a spin-half particle; merely that the formalism of spin-half symmetry is convenient for describing it. 
We define the vector $\boldsymbol\alpha = (\alpha_1, \alpha_2)$ and the matrix 
$\widehat{\sigma}_z = (
\begin{smallmatrix}
1&0\\0&-1
\end{smallmatrix}
)$ so that \eqref{eq:cos-theta} becomes
\begin{equation}
	\sigma_z ~~=~~ \frac{\boldsymbol\alpha^* .~\widehat{\sigma}_z \boldsymbol\alpha}
			   {\boldsymbol\alpha^* . \boldsymbol\alpha}
\label{eq:sigma-z}
\end{equation}
This generalises to an axis-independent form as follows (see for example~\cite{rae2008quantum} for further detail). 
The state $\boldsymbol\alpha
= (1,1)/\sqrt{2}$ has $(\theta, \varphi) = (\frac{\pi}2, 0)$, from
\eqref{eq:bloch-state}, and its `spin projections' $(\sigma_x,
\sigma_y, \sigma_z)$, defined as in \eqref{eq:sigma-z} with the conventional Pauli matrices, are $(1,0,0)$,
which is in the same direction.  The same is true for $y$ and $z$, and
so if $(\theta, \varphi)$ is in the direction ${\bf a}$ then its spin
projection on to ${\bf b}$ is
\begin{equation}
	\sigma_b
	~~=~~ \frac{
				\boldsymbol\alpha^* . ({\bf b}.\widehat{\boldsymbol\sigma}) \boldsymbol\alpha
				}
				{
				\boldsymbol\alpha^* . \boldsymbol\alpha
				}
				~~=~~ {\bf a}.{\bf b}
\label{eq:spin-dot-product}
\end{equation}
where ${\bf b}.\widehat{\boldsymbol\sigma}$ means $\sum b_i
\widehat\sigma_i$ and $|a| = |b| = 1$. 

We have seen that the energy and angular momentum of a quasiparticle are reduced by having an anticorrelated partner. 
If one quasiparticle, $\alpha$, is in the Bloch state \eqref{eq:bloch-state} and its partner, $\alpha'$, is in the mirror state, then defining $\xi_{ud} = \xi_{z+} - \xi'_{z-}$ and $\xi_{du} = \xi_{z-} - \xi'_{z+}$, the joint state is
\begin{equation}
	\xi_{\alpha \alpha'}~~ =~~ e^{i(S-\frac12 \varphi)} (\cos \tfrac{1}{2} \theta~ \xi_{ud} + e^{i \varphi} \sin \tfrac{1}{2} \theta~ \xi_{du})
\label{eq:bloch-ud}
\end{equation}

\begin{figure}[htb]
	\centering
		\includegraphics[width=.4\textwidth]{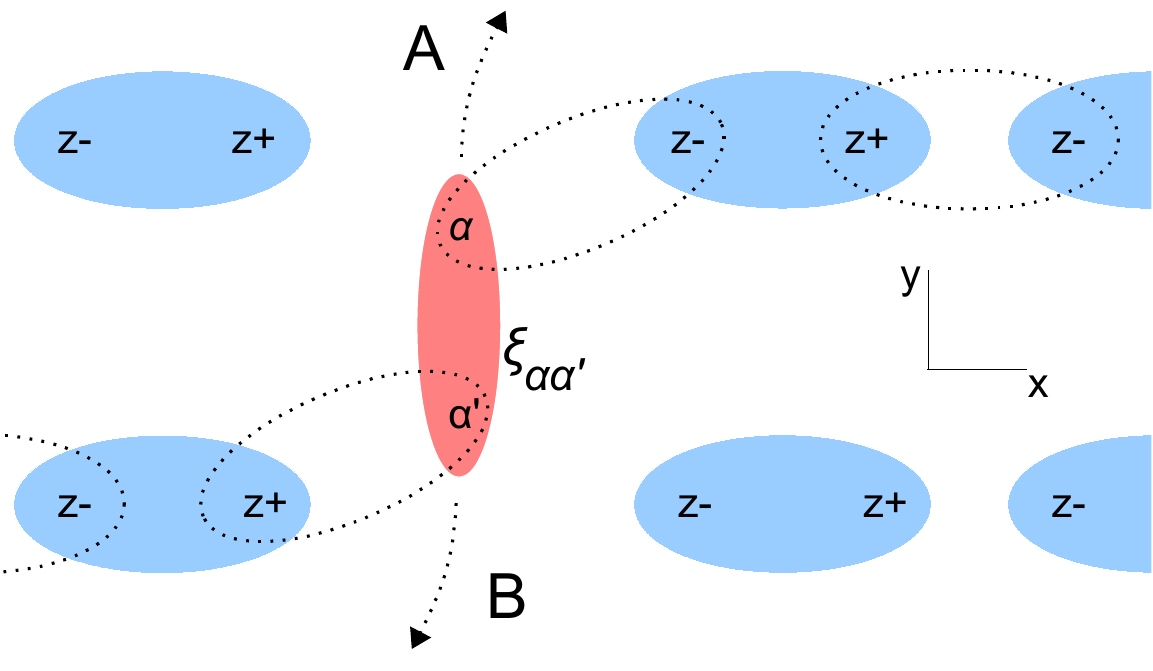}
	\caption{\em A pair of quasiparticles in the state $\xi_{\alpha \alpha'}$ separate and approach the local spins.}
	\label{fig:mead-transition}
\end{figure}

\ifheaders
	\section*{Spin correlation}
\else
	{\bf Spin correlation.}
\fi
In figure \ref{fig:mead-transition}, two quasiparticles are created
together in the state $\xi_{\alpha \alpha'}$ in \eqref{eq:bloch-ud}. They separate and approach regions $A$ and $B$ where the local spin directions are aligned with the $z$ axis (out of the paper) as shown. 
As they approach, new states become possible. 
The states drawn dotted are coupled to $\xi_{\alpha \alpha'}$ through the component $\xi_{ud}$ from the first term in \eqref{eq:bloch-ud}, and states in the opposite horizontal direction (not drawn) couple through $\xi_{du}$ in the second. 

The linearised approximation to Euler's equation is inadequate for describing the coupling, since linear solutions are additive and do not interact. 
The coupling arises from quadratic terms such as the Bernoulli pressure $P_B \propto u^2 = (c \Delta \rho/\rho_o)^2$. 
This oscillates at double frequency since it is quadratic in $\Delta \rho$, so it excites the new state parametrically. 
(See Couder's experiments on parametrically driven fluid motion~\cite{couder2006single,eddi2009unpredictable}.)
Initially, a parametrically excited state grows exponentially, $A \propto e^{\nu t}$, and in later stages, from the symmetry, the donating state $\xi_{\alpha \alpha'}$ diminishes exponentially at the same rate $e^{-\nu t}$ until it loses coherence. 
The transition cannot easily be reversed, since this would require a new and precise alignment of phase and frequency. 
Mead compares similar resonant transitions to quantum jumps~\cite{mead2000collective}.

The paths will be deflected as shown in figure \ref{fig:mead-transition} due to the attraction in a bound pair discussed above. 
Counting a deflection to the right as $+1$ and to the left
as $-1$, the product of the deflections is $-1$ in both cases.
In Bell's thought experiment the details of the apparatus are different, but the deflections are likewise always opposed when the Stern-Gerlach magnets are parallel~\cite{bell1964einstein}.

We wish to calculate the mean product of the deflections (the `correlation') when the spins near $A$ have been rotated through angle $\theta$ about the $y$ axis. We will continue to analyse the pair as a superposition of states that are aligned with the $z$ axis as in \eqref{eq:bloch-ud}; this does not lose generality because it spans the relevant states. 

We begin with the component $\xi_{ud}$, for which the deflection at $B$ is as drawn, counting as $-1$. The probability of this is $P_{b-} = \frac12$. 

The deflections at $A$ can be calculated in new axes at angle $\theta$, so that the local spins are aligned with the new $z$ axis and the component is described in the form \eqref{eq:bloch-ud}. 
The transitions in the $+\theta$ direction are coupled through the first term in \eqref{eq:bloch-ud}, and they are driven by quadratic terms such as the Bernoulli pressure, so that $\nu \propto \cos^2 \frac12 \theta$. 
The rate-limiting step is in the very early stages of the transition, when phase locking must be established against competing noise factors. 
No phase locking will take place if the driving excitation is zero, and so the rate, and hence the transition probability, is proportional to $\nu$ to first order (an example of a first order term can be seen when the disruptions occur at random times). Thus, the transition probability is $P_{a+|b-} = \cos^2 \frac12 \theta$. In the other direction, $P_{a-|b-} = \sin^2 \frac12 \theta$.
 
We will define $P_{+-} = P_{a+|b-} P_{b-} = \frac12 \cos^2 \frac12 \theta$ and $P_{--} = P_{a-|b-} P_{b-} = \frac12 \sin^2 \frac12 \theta$. 
Similarly, from $\xi_{du}$, $P_{++} = \frac12 \sin^2 \frac12 \theta$ and $P_{-+} = \frac12 \cos^2 \frac12 \theta$, and the required correlation is
\[
P_{++} ~+ ~P_{--} ~-~ P_{-+} ~-~ P_{+-} ~~=~~ -\cos \theta
\]

If the two directions are
$\bf a$ and $\bf b$, this correlation is 
$-{\bf a}.{\bf b}$. 
It is independent of the axes chosen for analysis, as for the spin projection in \eqref{eq:spin-dot-product}.
It is the same as the prediction for quantum-mechanical particles when the Stern-Gerlach magnets are at angle $\theta$ to one another.

Bell showed this correlation to be inconsistent with his `requirement
of locality,' as he posited it for point-like classical particles, and
tests have shown that quantum phenomena do indeed violate his locality
requirement.  But so do collective phenomena in classical fluid
mechanics. See~\cite{brady2013incompressible,anderson2013quantum} for
discussion of time-varying Bell experiments and the implications for
quantum computing.

\ifheaders
	\section*{Conclusion}
\else
	{\bf Conclusion.}
\fi
We have shown that the quasiparticle solutions to Euler's equation for a compressible inviscid fluid are correlated in precisely the same way as the quantum-mechanical particles discussed in Bell's original paper. The violation of Bell's inequality occurs because the fluid motion is correlated over a large distance.

We conclude that Bell's analysis does not exclude the possibility of purely local interactions underlying and explaining quantum mechanics. 

\ifprl
	\begin{acknowledgments}
\else
	\pagebreak
\fi

We thank Robin Ball, Graziano Brady, Joy Christian, Boris Groisman, Gerhard Gr\"ossing, Basil Hiley, Ruth Kastner, Michael McIntyre, Carver Mead, Hrvoje Nikolic, David Turban and Louis Vervoort for comment and discussion.

\ifprl
	\end{acknowledgments}
\else
\fi

\end{document}